\documentclass[aps,twocolumn,nofootinbib]{revtex4-1}
\usepackage{amsmath,graphicx}
\newcommand{\avgev}[1]{\left\langle{#1}\right\rangle}
\newcommand{\avgevvn}[1]{\left\langle{#1}\right\rangle_{|v_n}}
\newcommand{\avgvn}[1]{\left\langle{#1}\right\rangle_{v_n}}

\newcommand{\Qf}[2]{\frac{Q_{#1}^{#2}}{|Q_{#1}^{#2}|}}
\newcommand{\Qfs}[2]{\frac{Q_{#1}^{*#2}}{|Q_{#1}^{#2}|}}
\begin{document}
\title{Eliminating experimental bias in anisotropic-flow measurements \\of high-energy nuclear collisions}
\author{Matthew Luzum, Jean-Yves Ollitrault}
\affiliation{
CNRS, URA2306, IPhT, Institut de physique theorique de Saclay, F-91191
Gif-sur-Yvette, France}
\date{\today}

\begin{abstract}
We argue that the traditional event-plane method, which is still
widely used to analyze anisotropic flow in ultrarelativistic heavy-ion
collisions, should be abandoned because flow fluctuations introduce an
uncontrolled bias in the measurement. 
Instead, one should use an alternative, such as the scalar-product method or cumulant method, which always measures an unambiguous property of the underlying anisotropic flow and therefore eliminates this bias, and does so without any disadvantages.   
It is known that this correction is important for precision
comparisons of traditional $v_n$ measurements requiring better than a
few percent accuracy.  However, we show that it is absolutely essential for
correlations between different harmonics, such as those that have been recently measured by the ATLAS Collaboration, which can differ from the nominally-measured quantity by a factor two or more.  We also describe how, using the corrected analysis method, the
information from different subevents can be combined in order 
to optimize the precision of analyses. 
\end{abstract}
\maketitle
\section{Introduction}

Anisotropic flow $v_n$, in particular elliptic flow $v_2$, is one of the
most important observables of ultrarelativistic heavy-ion 
collisions~\cite{Ackermann:2000tr,Back:2002gz,Alt:2003ab,Adler:2003kt,Aamodt:2010pa,Chatrchyan:2012wg,ATLAS:2012at,Adamova:2012md}. 
It is therefore important to devise the best possible methods to analyze it. 
A good measurement should  be reproducible; in particular, it should be done in such a way that one can easily compare results from 
different experiments, using different detectors. 
For the sake of comparison with theory, an ideal measurement is a well-defined quantity that corresponds to a generic property of the system, closely related to an interesting theoretical concept.
Any such measurement is superior to one that lacks these traits.

In this paper, we argue that the traditional event-plane method~\cite{Poskanzer:1998yz}, which is still used by most of the 
major experimental collaborations~\cite{ATLAS:2012at,Adamova:2012md,Adare:2011tg,Chatrchyan:2012ta,Abelev:2012di}
does not satisfy any of the above requirements defining a good measurement. 
When the method was first developed, it was assumed that event-by-event fluctuations in $v_n$ were negligible.
However, it is now known that $v_n$ fluctuates significantly within a class of events~\cite{Miller:2003kd,Alver:2006wh,Alver:2010gr}. 
As a result, an event-plane measurement of $v_n$ yields an ambiguous measure lying somewhere between 
the event-averaged mean value $\langle v_n\rangle$ and the root-mean-square value $\langle v_n^2\rangle^{1/2}$~\cite{Alver:2008zza,Ollitrault:2009ie}.
Where exactly depends on the ``resolution'', which strongly depends on the 
experimental setup. 
This means that, e.g., the same Pb-Pb collisions at the LHC yield a systematically different value of $v_n$ depending on whether 
they are analyzed by ALICE, CMS or ATLAS, and that comparisons to earlier measurements from lower energy collisions at RHIC are ambiguous. 
The difference is typically of a few percent for elliptic flow~\cite{Ollitrault:2009ie}, but is larger for higher harmonics like
triangular flow $v_3$~\cite{Alver:2010gr}, which is solely due to fluctuations. 

Fortunately, there exist alternative analysis methods that do not suffer from this ambiguity, such as 
the two-particle cumulant method~\cite{Borghini:2001vi,Bilandzic:2010jr} or a slight variant of the 
traditional event-plane method called the scalar-product method introduced by 
the STAR collaboration~\cite{Adler:2002pu}.
These make for a superior measurement because they consistently yield 
the rms value of $v_n$, while introducing no disadvantage compared to the traditional event-plane measurements.  Supplemented by higher order cumulant measurements, these make traditional event-plane measurements completely redundant, as they contain no independent information about the underlying flow.

Traditional event-plane analysis methods have also been used to measure higher-order, mixed-harmonic correlations~\cite{Adams:2003zg,Adare:2010ux,Jia:2012sa}.  These measurements also suffer from similar biases as traditional event-plane $v_n$ measurements, but in this case they are much larger~\cite{Gombeaud:2009ye}, causing significant problems even for qualitative comparisons.  Fortunately, these problems can be repaired with a few simple changes.

In Secs.~\ref{s:flowvector} and~\ref{s:subevents},
we recall the principle of the traditional event-plane method and of its variant, the scalar-product method. 
In Sec.~\ref{s:flowfluctuations},
we show that the former yields an ambiguous measurement in the presence of flow fluctuations, while the 
latter yields an unambiguous result. 
These results are already known to experts~\cite{Alver:2008zza}. 
For purposes of illustration, we make several simplifications compared to what is typically done in an experiment, 
and in Sec.~\ref{s:details} discuss these complications and why they do not change the results.
We also describe in Sec.~\ref{s:combining} a new, systematic way of 
improving the accuracy of the scalar-product method 
by combining reference flows from ``subevents''.  
In Sec.~\ref{s:correlations}, the discussion is extended to the analysis of mixed correlations, 
where we show for the first time that the ambiguity resulting from flow fluctuations is much 
larger, and the proposed corrected procedure is therefore critical.

\section{Measuring $v_n$}
\label{s:spmethod}

The picture underlying anisotropic flow measurements is that, if the system exhibits strong collective behavior (i.e., ``flow''), particles 
in each event are emitted independently according to an underlying one-particle probability distribution~\cite{Luzum:2011mm}
\begin{equation}
\label{prob}
\frac{2\pi}{N}\frac{dN}{d\phi}=1+2\sum_{n=1}^{\infty}v_n \cos n(\phi-\Phi_n),
\end{equation}
where $\phi$ is the azimuthal direction of an emitted particle, $v_n$ is the amplitude of anisotropic flow in the $n$th harmonic, and $\Phi_n$ the corresponding reference angle.  Further assumptions underlying most analyses are that $\Phi_n$ represents a global angle that depends little on pseudorapidity and transverse momentum,  that event-by-event fluctuations of $v_n$ also depend little on pseudorapidity and transverse momentum, and that multiplicity fluctuations within a centrality bin are negligible.

Experiments typically detect anywhere from a few hundred particles to at most a few thousand in a given collision event, while the anisotropy coefficients $v_n$ are typically a few percent or less.  A precise reconstruction of the underlying probability distribution is therefore impossible in a single event.  Information about the underlying probability distribution can only be extracted from (azimuthally symmetric) correlations between outgoing particles, averaged over a large ensemble of events~\cite{Wang:1991qh}.  

\subsection{The flow vector}
\label{s:flowvector}

The various experimental estimates of $v_n$ have compact expressions
in terms of the flow vector $\vec Q_n$ in harmonic $n$. 
For a given set of $N$ particles belonging to the same event, one defines $\vec Q_n\equiv (|Q_n|\cos(n\Psi_n),|Q_n|\sin(n\Psi_n))$ by~\cite{Poskanzer:1998yz}
\begin{eqnarray} 
\label{defqvector}
|Q_n|\cos(n\Psi_n)&=& \frac 1 N \sum_j \cos(n\phi_j)\cr
|Q_n|\sin(n\Psi_n)&=&\frac 1 N \sum_j \sin(n\phi_j),
\end{eqnarray} 
where the sum runs through the set of $N$ particles with respective azimuthal angles $\phi_j$.
The flow vector \eqref{defqvector} can also be written in complex form
\begin{equation}
\label{defqcomplex}
Q_n = |Q_n| e^{in\Psi_n}\equiv\frac1 N \sum_j e^{in\phi_j}.
\end{equation}

The original idea behind the event-plane method is that the direction
$\Psi_n$ of the flow vector in a reference detector provides an
estimate of the corresponding 
angle $\Phi_n$ in the underlying probability
distribution~\cite{Danielewicz:1985hn}. Because a 
finite sample of particles is used, statistical fluctuations cause $\Psi_n$ to differ from $\Phi_n$. 
This dispersion is characterized by the ``resolution'', defined as:
\begin{equation}
\label{R0}
  R\equiv\avgev{ e^{in(\Psi_n-\Phi_n)}} =
  \avgev{\Qf{n}{}e^{-in\Phi_n}}, 
 \end{equation}
where angular brackets denote an average over events. 
Note that this average is real by parity symmetry, except for
irrelevant statistical fluctuations. 

When the method was developed, it was assumed that dynamical fluctuations 
of the underlying probability distribution are negligible, so that
$v_n$ is the same for all events. We then rewrite Eq.~(\ref{R0}) as 
\begin{equation}
\label{R}
 R(v_n)\equiv\avgevvn{ e^{in(\Psi_n-\Phi_n)}} = \avgevvn{\Qf{n}{}e^{-in\Phi_n}}.
 \end{equation}
where $\avgevvn{\ldots}$ indicates an average over a large number of events with the same underlying $v_n$,
The dependence of $R$ on $v_n$ can be easily understood if one sees 
Eq.~(\ref{defqvector}) as a directed random walk: particles are
emitted randomly, but the underlying probability distribution
(\ref{prob}) is anisotropic. 
The resolution $R$ thus depends on the relative magnitude of the anisotropy
$v_n$ to the statistical dispersion $1/\sqrt{N}$.
In the limit $v_n\gg 1/\sqrt{N}$ 
(infinite number of particles), one can exactly reconstruct 
the underlying event 
plane so that $\Phi_n = \Psi_n$, and 
\begin{equation}
\label{Rhigh}
R(v_n) \xrightarrow[\text{high res.}]{} 1.
\end{equation}
Conversely, when the resolution is low ($v_n\sqrt{N}\ll 1$),
\begin{equation}
\label{Rlow}
R(v_n) \xrightarrow[\text{low res.}]{} 
{k v_n},
\end{equation}
where $k$ is independent of $v_n$ and scales as $k\sim\sqrt{N}$.

In the general case, the value is somewhere in between these limits.
Analytic formulas can be obtained for $R(v_n)$ under rather general
assumptions ($N\gg 1$, $v_n\ll 1$, $v_{2n}\ll
1$)~\cite{Ollitrault:1997di}, but they are of little practical use
except when combining subevents. 

This nonlinear dependence of the resolution on the underlying flow is
the origin of the difficulties of the event-plane method, which arise
when flow fluctuations are considered. 
A simpler quantity 
is the projection of the flow vector onto the underlying 
direction $\Phi_n$, which directly gives the underlying flow:
\begin{equation}
\label{vn}
\avgevvn{ Q_ne^{-in\Phi_n}} = v_n,
\end{equation}
where we have used Eq.~(\ref{prob}).

\subsection{Correlating subevents}
\label{s:subevents}

Historically, the most common way to access information about the coefficients $v_n$ has been so-called event-plane measurements.
In these analyses,
$v_n$ is extracted from correlations between particles of interest
(e.g., identified particles in a narrow transverse momentum window) and a 
different set of particles in a reference detector
(e.g., unidentified particles in a large $p_T$ range), typically
separated by a gap in pseudorapidity~\cite{Adler:2003kt,Alver:2010rt}.  

The event-plane method thus correlates the flow vector of particles of 
interest, denoted by $Q_{n}$, with the direction of the flow vector
in a reference detector $A$, denoted by $Q_{nA}$. 
Assuming that the only correlation between $Q_{n}$ and $Q_{nA}$ is
that resulting from their  
correlation with the direction of anisotropic flow $\Phi_n$, 
this correlation factorizes:
\begin{eqnarray}
\label{factorization}
\avgevvn{Q_{n}\Qfs{nA}{}}&=&
\avgevvn{Q_{n}e^{-in\Phi_n}}
\avgevvn{\Qf{nA}{}e^{-in\Phi_n}}^*\cr
&=&v_n R(v_{nA}),
\end{eqnarray}
where we have used Eqs.~\eqref{vn} and \eqref{R}, and $v_{nA}$
denotes the value of $v_n$ in the reference detector $A$. 
The resolution $R(v_{nA})$ is estimated by correlating $A$ with
additional separate reference detectors; 
Each reference detector is called a ``subevent''.
In the simplest 
case of two identical subevents ($A$, $B$) located
symmetrically around midrapidity, making use of the factorization
hypothesis: 
\begin{eqnarray}
\label{factorization2}
\avgevvn{\Qf{nA}{}\Qfs{nB}{}}&=&
\avgevvn{\Qf{nA}{}e^{-in\Phi_n}}
\avgevvn{\Qf{nB}{}e^{-in\Phi_n}}^*\cr
&=&
\left|\avgevvn{\Qf{nA}{}e^{-in\Phi_n}}\right|^2\cr
&=&R(v_{nA})^2,
\end{eqnarray}
where we have again used Eq.~\eqref{R}. 
The event-plane measurement is thus defined as
\begin{equation}
\label{defvnep}
v_n\{EP\}\equiv \frac{\avgev{Q_{n} \Qfs{nA}{}}}{\sqrt{\avgev{\Qf{nA}{}\Qfs{nB}{}}}},
\end{equation}
Since it is impossible to select events according to the underlying probability distribution,
the measurement is taken with averages over all events in a centrality class, denoted by 
unadorned angular brackets.
In the absence of flow fluctuations (i.e., if the underlying $v_n$ is
the same in every event), 
Eqs.~\eqref{factorization} and~\eqref{factorization2} 
ensure that $v_n\{EP\}$ coincides with the underlying $v_n$, up to
experimental errors.  

A slight variant of the event-plane method consists in removing the factors of $|Q_n|$ before taking the average in 
the numerator and denominator of Eq.~\eqref{defvnep}.
\begin{equation}
\label{defvnsp}
v_n\{SP\}\equiv \frac{\avgev{Q_{n} Q_{nA}^*}}{\sqrt{\avgev{Q_{nA}Q_{nB}^*}}},
\end{equation}
The numerator of Eq.~(\ref{defvnsp}) involves a scalar product of two
vectors $\vec Q_{n}\cdot \vec Q_{nA}$,  
and so is referred to as the scalar-product method~\cite{Adler:2002pu}.
In the absence of flow fluctuations, factorization again applies:
\begin{eqnarray}
\label{factorization3}
\avgevvn{Q_{n}Q_{nA}^*}&=&
\avgevvn{Q_{n}e^{-in\Phi_n}}
\avgevvn{Q_{nA}e^{-in\Phi_n}}^*\cr
&=&v_nv_{nA}
\end{eqnarray}
and similarly
\begin{equation}
\label{factorization4}
\avgevvn{Q_{nA}Q_{nB}^*}=v_{nA}^2.
\end{equation}
Therefore $v_n\{SP\}$ also coincides with the underlying $v_n$ in the
absence of flow fluctuations.

\subsection{Flow fluctuations}
\label{s:flowfluctuations}

However, it is now known that $v_n$
fluctuates significantly from event to event~\cite{Alver:2006wh}. 
To make it clear what is really measured in this case,
we evaluate event averages in two steps, by first 
averaging over events with the same $v_n$, then combining
bins of $v_n$. 
\begin{equation}
\label{twosteps}
\avgev{\ldots} \equiv \avgvn{\avgevvn{\ldots}}.
\end{equation}
We apply this decomposition to Eq.~\eqref{defvnep} and evaluate the
inner average using Eqs.~\eqref{factorization} and
\eqref{factorization2}:
\begin{equation}
v_n\{EP\} 
= \frac{\avgvn{v_n R(v_{nA})}}{\sqrt{\avgvn{R(v_{nA})^2}}},
\end{equation}
Note that $\langle R(v_{nA})^2\rangle\not=\langle R(v_{nA})\rangle^2$, 
so that the resolution correction
is no longer  a simple projection of the measured event plane $\Psi_n$
onto the ``true'' event plane $\Phi_n$.

In the limit of perfect resolution (i.e., the number of particles used in the reference detector goes to infinity), $R(v_{nA})\simeq 1$ and $v_n\{EP\}$ does indeed measure the event-averaged mean $v_n$ from Eq.~\eqref{prob}, 
\begin{equation}
\label{vnhigh}
v_n\{EP\} \xrightarrow[\text{high res.}]{} \avgev{ v_n} .
\end{equation}
In reality, the resolution is not perfect, and the result is usually closer to the low resolution limit~\cite{Alver:2008zza}. 
In this limit, 
$R(v_{nA})\propto v_{nA}$. 
Assuming that flow fluctuations are global, i.e., $v_n/v_{nA}$ does not fluctuate significantly,   
the event-plane measurement thus yields a root-mean-square value, 
\begin{equation}
\label{vnlow}
v_n\{EP\} \xrightarrow[\text{low res.}]{} 
\sqrt{\avgev{ v_n^2 }} .
\end{equation}
\begin{figure}
\includegraphics[width=\linewidth]{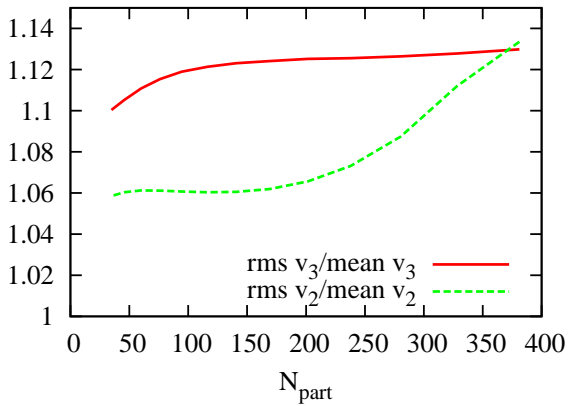}
\caption{(Color online) Ratio of the rms $v_n$, corresponding to the low-resolution limit of Eq.~(\ref{defvnep}) integrated over $p_T$, to the mean $v_n$, corresponding to the high-resolution limit, as a function of the number of participant nucleons (binned in 5\% centrality intervals), for a Pb-Pb collision at 2.76 TeV per nucleon pair. We model flow fluctuations by assuming $v_n\propto\varepsilon_n$ in each event, where $\varepsilon_n$ is the anisotropy of the initial distribution in the corresponding harmonic, defined as in Ref.~\cite{Gardim:2011xv}. The value of  $\varepsilon_n$ in each event is obtained from the Phobos Monte Carlo Glauber model~\cite{Alver:2008aq}. For $v_3$, the ratio is close to $2/\sqrt{\pi}\simeq 1.13$, corresponding to Gaussian fluctuations~\cite{Voloshin:2007pc}. }
\label{fig:ratios}
\end{figure}
In general, the event-plane method yields a result which may lie
anywhere between the two limits \eqref{vnhigh} and \eqref{vnlow}. The
exact measured quantity depends the detector acceptance---even after
the resolution correction is applied---so that the measurement is ambiguous. 

The difference between the two limits 
ranges from 6 to 13\%, as illustrated in Fig.~\ref{fig:ratios}.  As the field enters an era of precision physics,  comparisons of different experimental analyses~\cite{Ollitrault:2009ie} and of theory to experiment~\cite{Gardim:2012yp} are significantly complicated by this dependence on
analysis details.  For example, the large detector acceptance and high multiplicities of recent measurements at the LHC allow for a larger resolution than was available at RHIC.  The natural choice (and what has been done) is to take advantage of this extra resolution to minimize the statistical uncertainty of the measurement.  The penalty, however, is that comparisons to earlier measurements become ambiguous, and one has difficulty determining from these measurements how much the flow changes, e.g., with collision energy.   

Theoretical comparisons are hampered for the same reason.  Theorists have access to more information than experiments --- their `detectors' have perfect efficiency and suffer no dead spots or holes in coverage; i.e., they can see all particles that are produced.  However, if they succumb to the temptation to make use of this extra information to improve statistics (or they simply lack detailed information about detector performance to appropriately degrade the data) comparisons to measurements can become ambiguous and unreliable --- even if they otherwise follow the exact same procedures as the experiment~\cite{Holopainen:2010gz,Petersen:2011sb,Pang:2012he}.

Fortunately, this ambiguity can be removed by using the scalar-product  method. 
Although it was originally introduced for unrelated reasons, it gives a well-defined measurement in the presence of flow fluctuations~\cite{Ollitrault:2009ie},
despite differing only trivially from the traditional analysis.
When $v_n$ fluctuates, we again apply the decomposition
\eqref{twosteps} to Eq.~\eqref{defvnsp} and evaluate the numerator and
the denominator using Eqs.~\eqref{factorization3} and \eqref{factorization4}:
\begin{equation}
v_n\{SP\} 
= \frac{\avgvn{v_n v_{nA}}}{\sqrt{\avgvn{v_{nA}^2}}}
=\sqrt{\avgev{v_n^2}},
\end{equation}
where we have again used the hypothesis that $v_n/v_{nA}$ does not
fluctuate significantly. 
Therefore, the scalar-product method always yields the root-mean-square $v_n$, regardless of the details of the analysis, and makes for a superior measurement. 

One can also choose to measure a two-particle correlation~\cite{Aamodt:2011by}, 
which is essentially equivalent to a scalar-product analysis, and also always measures 
this same unambiguous root-mean-square value~\cite{Miller:2003kd}.  It is important to 
emphasize, however, that even when an event-plane-type analysis is preferred for practical 
reasons, only a trivial change to the standard analysis is required in order to obtain a 
well-defined observable.

If the resolution is large enough, higher-order cumulants~\cite{Borghini:2001vi,Bilandzic:2010jr} can be analyzed.  These then yield an unambiguous measurement of higher-order, even moments of the distribution of $v_n$, i.e. $\langle (v_n)^{2k}\rangle$~\cite{Miller:2003kd}. 
Since a standard event-plane method measures a quantity that is different from any of these 
individual observables, one might naively think it useful to perform both analyses in order to 
obtain the maximum amount of independent information.  However,
it has been proven~\cite{Ollitrault:2009ie} that the event-plane method, in addition to being ambiguous, contains no independent information beyond that contained in the first two cumulants,  $v_n\{2\}$ and $v_n\{4\}$ (regardless of whether nonflow correlations are present in the system).  So there is no reason to use a standard event-plane method in any future analysis.

\subsection{Practical details}
\label{s:details}

Actual measurements often differ slightly from the above idealized discussion.   The flow vector~\eqref{defqcomplex} is typically defined~\cite{Poskanzer:1998yz} without the normalization factor $N$.  Similarly, the numerators in Eqs.~\eqref{defvnep} and~\eqref{defvnsp} are calculated as a sum over the particles of interest in all events, before normalizing by the total number of particles after the sum over events.  Thus, there are weighting factors of $N$ and $dN/dp_T$, respectively, inside the event averages, which become important only if large enough centrality bins are used such that multiplicity and spectrum fluctuations are significant.  

Second, the flow vector used to estimate the event plane is often calculated as a weighted average of particles.  
In particular, if it is measured using a calorimeter, the particles are weighted according to their energy, while
a $p_T$-dependent weight is often used to increase the resolution or to remove non-flow 
correlations~\cite{Luzum:2010fb}.  In this case $v_{nA}$ above refers
to a weighted average.

Third, in the case where identical subevents are not available, the
analysis can be done with three arbitrary subevents $A$, $B$, and
$C$~\cite{Poskanzer:1998yz}. One then does the following replacement
\begin{equation}
\avgev{Q_{nA}Q_{nB}^*}\rightarrow\frac{\avgev{Q_{nA}Q_{nB}^*}\avgev{Q_{nA}Q_{nC}^*}}{\avgev{Q_{nB}Q_{nC}^*}}
\end{equation}
in the denominator of Eq.~\eqref{defvnsp}, and a similar modification
in the resolution correction of Eq.~\eqref{defvnep}. 

Fourth, we have used a complex notation throughout this
section. However, all quantities are real after averaging over events,
except for irrelevant experimental errors. Therefore $\avgev{\ldots}$
should be replaced by $\Re\avgev{\ldots}$ everywhere. 

Finally, if any of the flow vectors in the numerator of Eqs.~\eqref{defvnep} and~\eqref{defvnsp} 
are calculated from the same or overlapping sets of particles, care must be made to remove self-correlations~\cite{Danielewicz:1985hn}.

\subsection{Combining subevents}
\label{s:combining}

We end this section with a brief description of how the precision of the scalar-product method can be improved beyond previous implementations. 
Since at least two reference detectors $A$ and $B$ are needed in order
to carry out the flow analysis, it is tempting to combine the information from both detectors into a single measurement with reduced statistical uncertainty. In the standard event-plane method, this can be done for identical 
subevents at the expense of additional hypotheses and algebraic complications \cite{Poskanzer:1998yz,Ollitrault:1997di}, 
but there is no known way of recombining non-identical subevents, as in the three-subevent method. 

With the scalar-product method, combining measurements from two reference detectors $A$ and $B$ is straightforward, even if they are not identical. 
One simply measures $v_n$ independently with respect to $A$ and to $B$.
By linearity of the scalar product, combining both measurements (denoted by $v_{n}\{SP,A\}$ and $v_n\{SP,B\}$) amounts to taking 
a weighted average. We show in Appendix \ref{s:appendix} that the optimal weighting is 
\begin{equation}
\label{optweight}
v_n\{SP\}\equiv\frac{\chi_A^2 v_{n}\{SP,A\}+\chi_B^2 v_n\{SP,B\}}{\chi_A^2 +\chi_B^2 },
\end{equation}
where $\chi_A$ is the dimensionless resolution parameter. 
For identical subevents, $v_n$ is the mean of $v_{n}\{SP,A\}$ and
$v_n\{SP,B\}$ by symmetry:
\begin{equation}
v_n\{SP\}=\frac{v_{n}\{SP,A\}+v_n\{SP,B\}}{2}.
\end{equation}
For non-identical subevents, a third subevent $C$ is
required in order to determine the resolution parameter: $\chi_A$ is
given by 
\begin{equation}
\label{defchiA}
\frac{1}{\chi_A^2}+1\equiv
\frac{\langle (\vec Q_{nA})^2\rangle\langle\vec Q_{nB}\cdot\vec
  Q_{nC}\rangle}
{\langle\vec Q_{nA}\cdot\vec
  Q_{nB}\rangle\langle\vec Q_{nA}\cdot\vec Q_{nC}\rangle},
\end{equation}
and $\chi_B$ is given obtained through the substitution
$A\leftrightarrow B$ in Eq.~\eqref{defchiA}. 
Note that when detectors are combined, their resolution parameters
$\chi$ add in quadrature.  
Generalization of Eq.~\eqref{optweight} to three subevents $A$, $B$
and $C$ is straightforward. 

\section{Higher-order correlations}
\label{s:correlations}

Fourier harmonics of the azimuthal distribution are now measured up to $v_4$~\cite{Adare:2011tg} 
at RHIC and up to $v_5$~\cite{ALICE:2011ab,Chatrchyan:2012wg} and 
$v_6$~\cite{ATLAS:2012at} at LHC. 
Much additional information is contained in mixed correlations between different 
harmonics~\cite{Teaney:2010vd,Bhalerao:2011yg,Bhalerao:2011bp,Teaney:2012ke}, 
which give information about event plane angles $\Phi_n$ and will likely play a large part in upcoming measurements. 
Before rushing into analysis, it is important to define observables for good measurements, 
which will allow unambiguous comparison between different experiments, and with theory. 

The ATLAS collaboration has recently released preliminary measurements of an extensive set of correlations between 
event planes using traditional methods~\cite{Jia:2012sa}. For illustration, we consider the first of these correlations.  It is a correlation between the event planes $\Psi_4$ and 
$\Psi_2$, namely, $\langle\cos 4(\Psi_4-\Psi_2)\rangle$,  which is then divided by a resolution factor in order to unravel the correlation between the underlying reference directions $\Phi_4$ and $\Phi_2$. 
They thus write:
\begin{equation}
\label{atlascorrection}
\langle\cos 4(\Phi_4-\Phi_2)\rangle=\frac{\langle\cos 4(\Psi_4-\Psi_2)\rangle}
{\langle\cos 4(\Psi_4-\Phi_4)\rangle\langle\cos 4(\Psi_2-\Phi_2)\rangle}.
\end{equation}
This equation is based on the same hypotheses as the event-plane method, which fail if $v_2$ and $v_4$ fluctuate. Specifically, the numerator of the right-hand side does not factorize into 
$\langle\cos 4(\Psi_4-\Phi_4)\rangle\langle\cos 4(\Psi_2-\Phi_2)\rangle\langle\cos 4(\Phi_4-\Phi_2)\rangle$, 
and the resolution corrections estimated using standard procedures  do not actually correspond to 
$\langle\cos 4(\Psi_4-\Phi_4)\rangle$ and $\langle\cos 4(\Psi_2-\Phi_2)\rangle$.

Using the notations introduced in the previous section, and assuming
symmetric subevents $A$ and $B$ for simplicity, the observable
actually measured is
\begin{equation}
\label{atlas42}
\avgev{\cos 4 (\Phi_4 - \Phi_2)}\{EP\}
\equiv \frac
{\avgev{\Qf{4A}{}
\Qfs{2B}{2}}}
{\sqrt{\avgev{\Qf{4A}{}\Qfs{4B}{}}}\sqrt{\avgev{
\Qf{2B}{2}\Qfs{2A}{2}
}}}.
\end{equation}
In the presence of flow fluctuations, the average over events can be
evaluated in two steps following \eqref{twosteps}. 
In addition to $R_n$ given by Eq.~(\ref{R}), we
define a similar quantity $\mathcal{R}_n$
\begin{equation}
\label{R2}
\mathcal{R}_n\equiv \sqrt{\avgevvn{ e^{2in(\Psi_n-\Phi_n)}}} = \sqrt{\avgevvn{\left(\Qf
{n}{}e^{-in\Phi_n}\right)^2}}
 \end{equation}
 which has the same limits, Eqs.~\eqref{Rhigh} and \eqref{Rlow},  but can differ at an arbitrary resolution.
With this notation, 
\begin{equation}
\avgev{\cos 4 (\Phi_4 - \Phi_2)}\{EP\}
= \frac
{\avgvn{
R_4 \mathcal{R}_2^2 e^{ 4i (\Phi_4 - \Phi_2)}
}}
{
\sqrt{\avgvn{R_4^2}}
\sqrt{\avgvn{\mathcal{R}_2^4}}
}
\end{equation}

In the limit where the resolutions on both harmonics are perfect, the quantity measured by ATLAS is indeed the nominal value, corresponding to the left-hand side of Eq.~\eqref{atlascorrection}: 
\begin{equation}
\label{mixedhigh}
\cos 4(\Psi_4-\Psi_2)\{EP\}\xrightarrow[\text{high res.}]{}  \langle\cos 4(\Phi_4-\Phi_2)\rangle.
\end{equation}
In the limit of low resolution, however, repeating the same reasoning as in the previous section, 
what is actually measured is the quantity
\begin{equation}
\label{mixedlow}
\cos 4(\Psi_4-\Psi_2)\{EP\}\xrightarrow[\text{low res.}]{}
\frac{\langle v_{4A}v_{2A}^2\cos
  4(\Phi_4-\Phi_2)\rangle}{\sqrt{\langle v_{4A}^2\rangle\langle v_{2A}^4\rangle}}.
\end{equation}
As before, the measured value is generally between the two limits.\footnote{Note that the low-resolution limit \eqref{mixedlow} lies between $-1$ and $+1$ by construction, even though it does not reduce to a simple angular correlation.}
Similar expressions can be written for the other correlations in the large set measured by ATLAS. 
Unlike measurements of $v_n\{EP\}$, however, there is no known way to even estimate
where between these limits the result lies based only on the reported event-plane resolution.

Note that the same ambiguity plagues previous measurements of the similar ($p_T$-differential) quantity, $v_4$ with respect to the event-plane of 
elliptic flow $\Psi_2$~\cite{Adams:2003zg,Adare:2010ux}. 
Not only does the result depend on the resolution, it also depends on how the resolution correction is defined, as the seminal paper \cite{Poskanzer:1998yz} proposes two possible implementations, which turn out to be different in the presence of flow fluctuations. 

The high resolution limit of these measurements is the nominal quantity
\begin{equation}
\label{v4high}
v_4\{\Psi_2\}\xrightarrow[\text{high res.}]{}  \langle v_4\cos 4(\Phi_4-\Phi_2)\rangle. 
\end{equation}
The low-resolution limit depends on the actual implementation of the resolution correction. STAR~\cite{Adams:2003zg} uses Eq.~(11) of \cite{Poskanzer:1998yz}, which reduces to 
\begin{equation}
\label{v4lowSTAR}
v_4\{\Psi_2\}\xrightarrow[\text{low res.}]{}  \frac{\langle v_4
  v_{2A}^2\cos 4(\Phi_4-\Phi_2)\rangle}{\langle v_{2A}^2\rangle} 
\end{equation}
in the low-resolution limit, where again $v_{2A}$ is integrated over the reference detector. 
PHENIX uses the same implementation as ATLAS (Eq.~(14) or Eq.~(16) of \cite{Poskanzer:1998yz}) which yields, in the low-resolution limit,
\begin{equation}
\label{v4lowPHENIX}
v_4\{\Psi_2\}\xrightarrow[\text{low res.}]{} \frac{\langle v_4 v_{2A}^2\cos 4(\Phi_4-\Phi_2)\rangle}{\sqrt{\langle  v_{2A}^4\rangle}}.
\end{equation}

As recently 
demonstrated in Ref.~\cite{Gardim:2012yp} in full
event-by-event hydrodynamic calculations, this ambiguity of the event-plane method is a significant source of uncertainty when comparing experiment to theory. 
In fact, it should already have been apparent that the effect of
detector resolution in measurements such as these is much larger than
for the $v_n\{EP\}$ measurements described above.  
The difference between Eqs.~(\ref{v4lowSTAR}) and (\ref{v4lowPHENIX}) 
likely explains why the values of $v_4\{\Psi_2\}$ are systematically
higher for STAR than for PHENIX by some 10\%~\cite{Gombeaud:2009ye}. 
In Ref.~\cite{Adams:2003zg} the STAR collaboration measured
$v_4\{\Psi_2\}$, but in addition measured a 3-particle correlation
$v_4\{3\}$, which corresponds to the low-resolution limit of their event-plane measurement. 
The STAR collaboration
finds that $v_4\{3\}$ is larger than $v_4\{\Psi_2\}$ by 20\% for mid
central collisions. This difference is much larger than the difference
between, e.g., $v_2\{EP\}$ and the low resolution limit $v_2\{SP\}$ in
the same experiment (which is roughly 5\%).

In order to illustrate the possible full dependence on resolution, we use here a simplified calculation, though the results are consistent with more sophisticated calculations~\cite{Gardim:2012yp}.  It is known from event-by-event hydrodynamic calculations that $p_T$-integrated $v_2$, $v_4$, and the corresponding reference directions $\Phi_2$ and $\Phi_4$ can be accurately predicted in each event from the initial eccentricities according to~\cite{Gardim:2011xv}
\begin{eqnarray}
v_2 e^{2i\Phi_2}&=&a\varepsilon_2 e^{2i\Phi_{in,2}}\cr
v_4 e^{4i\Phi_4}&=&b\varepsilon_4 e^{4i\Phi_{in,4}}+c(\varepsilon_2 e^{2i\Phi_{in,2}})^2,
\end{eqnarray}
where $a$, $b$, $c$ are real coefficients, and $\varepsilon_n$ and $\Phi_{in, n}$ are the initial anisotropy in harmonic $n$ and its reference direction, as defined in Ref.~\cite{Gardim:2011xv}.

\begin{figure}
\includegraphics[width=\linewidth]{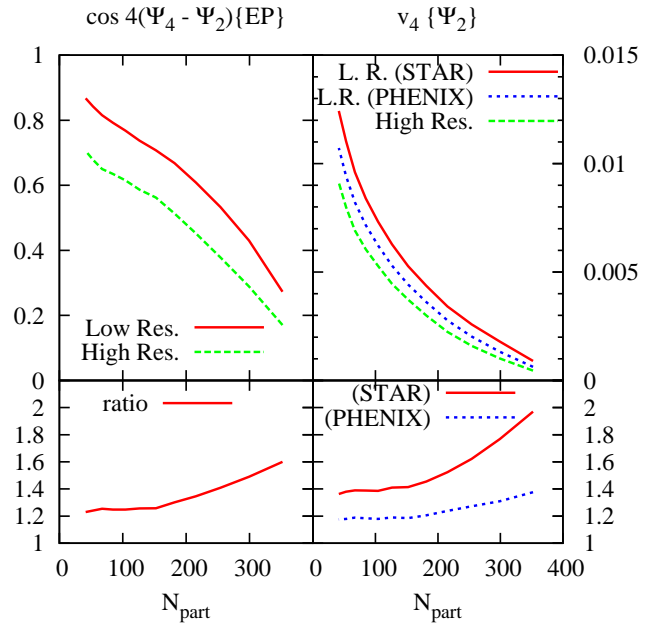}
\caption{(Color online) 
Top, left: low and high resolution limits of the correlation between $\Phi_2$ and $\Phi_4$, defined by Eqs.~\eqref{mixedlow} and \eqref{mixedhigh}, versus number of participants. 
Top, right: low and high resolution limits for integrated $v_4\{\Psi_2\}$, defined  by Eqs.~\eqref{v4lowSTAR},  \eqref{v4lowPHENIX} and \eqref{v4high}. Bottom panels: ratios of the low-resolution to high-resolution results. The nominally-measured values correspond to the high-resolution limit, which can differ from what is actually measured (usually closer to the low-resolution limit) by up to a factor 2.}
\label{fig:mixed}
\end{figure}

We thus compute $\varepsilon_n$ and $\Phi_{in,n}$ using the PHOBOS Monte Carlo Glauber~\cite{Alver:2008aq} and take the values of $b$ and $c$ from the viscous hydrodynamic calculation by Teaney and Yan~\cite{Teaney:2012ke}, for simplicity taking centrality-independent values of $b=0.018$ and $c = 0.07$~\footnote{Note that Teaney and Yan use cumulants of the initial distribution, which are linear combinations of our moments: with their notation, $b=w_4/{\cal C}_4$ and $c=w_{4(22)}/\varepsilon_2^2+3(w_4/{\cal C}_4)\langle r^2\rangle^2/\langle r^4\rangle$. We take $w_4/{\cal C}_4\simeq 0.018$ and  $w_{4(22)}/\varepsilon_2^2\simeq 0.04$ from their Fig.~2, right, and we estimate $\langle r^2\rangle^2/\langle r^4\rangle\simeq 0.6$ from a Glauber calculation.}.

The right-hand side of Eqs.~\eqref{mixedhigh} and \eqref{mixedlow} are displayed in 
Fig.~\ref{fig:mixed} (top, left), as well with their ratio (bottom, left), and similarly for Eqs.~\eqref{v4high},  \eqref{v4lowSTAR} and~\eqref{v4lowPHENIX} on the right side of the figure.  The ratio varies from 1.4 to more than 2 (in agreement with full event-by-event hydrodynamic calculations of $v_4\{\Psi_2\}$~\cite{Gardim:2012yp}). 
Comparing with Fig.~\ref{fig:ratios}, one sees that the ambiguity induced by the event-plane method is much worse for mixed correlations that for individual measurements of $v_n$. 

As a recent illustration of the problems that can arise from this, we point to the correlations calculated in Ref.~\cite{Qiu:2012uy}, corresponding to the high resolution limit of the recent preliminary ATLAS measurements --- i.e., the quantities that were nominally claimed as measured.  They are all clearly smaller in magnitude than the measurements, despite each observable having the correct sign and centrality dependence.  Naively, this seems to indicate that the theory needs to be changed somehow to accommodate this difference, but in fact the finite resolution of the experiment could in principle explain the entire discrepancy.  Further, while it is straightforward to calculate either the low or the high resolution limit, as done in that work, a calculation that reliably corresponds exactly to the measured quantity adds considerable difficulty. In fact, it is not possible without access to more information about the experimental details than is available.

Fortunately, the ambiguity of the event-plane method can easily be removed by using a straightforward generalization of the scalar-product method.  
The right-hand side of Eq.~\eqref{mixedlow} is given by
\begin{equation}
\label{mixedsp}
\frac{\langle v_{4A}v_{2A}^2\cos 4(\Phi_4-\Phi_2)\rangle}{\langle
  v_{2A}^2\rangle\sqrt{\langle v_{4A}^2\rangle}}=
\frac
{\avgev{Q_{4A}Q_{2B}^{*2}}}
{\sqrt{\avgev{Q_{4A}Q^*_{4B}}}\sqrt{\avgev{
Q_{2B}^2Q_{2A}^{*2}
}}},
\end{equation}
which we suggest as a replacement for Eq.~(\ref{atlas42}). 
In the same way as for individual measurements of $v_n$, this method
yields a result which is unambiguous and independent of the number of particles selected for analysis.
Note that the numerator is a sum over all triplets of particles of $\langle\cos(2\phi_1+2\phi_2-4\phi_3)\rangle$. The quantity in Eq.~\eqref{mixedsp} is thus identical to the type of scaled correlations introduced in~\cite{Bhalerao:2011yg}.

Generalization of the above discussion to $v_4\{\Psi_2\}$ and other mixed correlations~\cite{Jia:2012sa,Bhalerao:2011yg} is straightforward. In general, higher-order correlations involve higher moments of the distribution of anisotropic flow, and have therefore an increased sensitivity to flow fluctuations. 

Note that the particular correlation between $\Phi_2$ and $\Phi_4$ studied above can be contaminated by nonflow correlations; Therefore it is important that the flow vectors $Q_2$ and $Q_4$ are determined in regions separated by a pseudorapidity gap~\cite{Jia:2012sa}. Other mixed correlations, such as the correlation between $\Phi_2$ and $\Phi_3$, are not sensitive to nonflow correlations. The need for gaps in pseudorapidity must be considered on a case-by-case basis~\cite{Bhalerao:2011ry}.

\section{Conclusion}
We have explained that measurements of anisotropic flow using the event-plane method are ambiguous because the result depends on analysis details. This dependence, which is due to event-by-event flow fluctuations, is typically up to a few percent for $v_2$ and 10\% for $v_3$ and higher harmonics. 
For mixed correlations involving event planes from different harmonics, the dependence is much larger --- we have demonstrated that the result may vary by a factor 2 depending on analysis details. We have shown that these ambiguities of the event-plane method can be repaired with only minor modifications and without any additional complication or drawbacks. We therefore conclude that the traditional event-plane method should be abandoned in future analyses. 

\begin{acknowledgments}
ML is supported by the European Research Council under the Advanced Investigator Grant ERC-AD-267258.
We thank Instituto de Fisica, Sao Paulo, where this paper was written,
for hospitality, and FAPESP-CNRS for financial support (project 2011/51864-0)

\end{acknowledgments}

\appendix
\section{Combining reference flows}
\label{s:appendix}

In this Appendix, we show that Eq.~(\ref{optweight}) defines the optimal combination between 
anisotropic flows measured in detectors $A$ and $B$ using the
scalar-product method, in the sense that it minimizes the statistical
error.  

Eq.~\eqref{defvnsp} can be rewritten as 
\begin{equation}
\label{vnspbis}
v_n\{SP\}=\frac{\langle Q_n Q_{nA}^*\rangle}{\bar Q_{A}},
\end{equation}
(for simplicity, we drop the index $n$ from
now on) where $\bar Q_{A}$ is the resolution correction 
in subevent $A$, which is generally obtained using three subevents
$A$, $B$, $C$ pairwise separated by rapidity gaps: 
\begin{equation}
\label{defbarqna}
\bar Q_{A}\equiv\sqrt{\frac{\avgev{Q_{A}Q_{B}^*}\avgev{Q_{A}Q_{C}^*}}
{\avgev{Q_{B}Q_{C}^*}}}.
\end{equation}
 Assuming that $v_n\ll 1$ and that one measures $v_n$ in a narrow  
bin which contains at most one particle per event, 
so that $Q$ is essentially  a random number on the unit circle, 
and keeping in mind that one only keeps the real part in the numerator
of Eq.~(\ref{vnspbis}), 
a straightforward calculation shows that the statistical error on
$v_n\{SP\}$ is 
\begin{equation}
\label{staterr}
\delta v=\frac{1}{\sqrt{2N}}\frac{\sqrt{\langle |Q_A|^2\rangle}}{\bar
  Q_A},
\end{equation}
where $N$ is the number of particles in the bin.

If one measures $v_n$ in a detector that is separated in
pseudorapidity from both $A$ and $B$,  
the measurement can be done using either $A$ or $B$ as a reference.
We denote by $v\{SP,A\}$ and $v\{SP,B\}$ the corresponding estimates, where 
$v\{SP,A\}$ is given by Eq.~\eqref{vnspbis} and $v\{SP,B\}$ is given by
substituting $A\leftrightarrow B$ in Eqs.~\eqref{vnspbis} and
\eqref{defbarqna}. Note that by definition of $\bar Q_A$ and $\bar
Q_B$, 
\begin{equation}
\label{spproduct}
\langle Q_AQ_B^*\rangle=\bar Q_A\bar Q_B.
\end{equation} 

One can combine linearly the flow vectors and the resolution corrections
\begin{eqnarray}
\label{defQcomb}
Q_{\rm combined}&\equiv &Q_{A}+\lambda Q_{B}\cr
\bar Q_{\rm combined}&\equiv &\bar Q_A+\lambda \bar Q_B,
\end{eqnarray}
where $\lambda$ is arbitrary. 
The resulting combined measurement is 
\begin{equation}
\label{defcombined}
v\{SP\}\equiv\frac{ \langle QQ_{\rm combined}^*\rangle}{\bar
Q_{\rm combined}}
=\frac{\bar Q_A v\{SP,A\}+\lambda \bar Q_B v\{SP,B\}}{\bar Q_A+\lambda \bar Q_B}.
\end{equation}
Using Eq.~\eqref{staterr}, 
with $Q_A$ replaced by $Q_{\rm combined}$,
straightforward algebra shows that the value of $\lambda$ which
minimizes the statistical error on $v\{SP\}$ is 
\begin{equation}
\label{lambda}
\lambda=\frac
{\bar Q_B\langle |Q_A|^2\rangle-\bar Q_A\langle Q_AQ_B^*\rangle}
{\bar Q_A\langle |Q_B|^2\rangle-\bar Q_B\langle Q_AQ_B^*\rangle}.
\end{equation}
Using Eq.~\eqref{spproduct}, this can be rewritten 
\begin{equation}
\label{lambda2}
\lambda=\frac
{\bar Q_B(\langle |Q_A|^2\rangle-\bar Q_A^2)}
{\bar Q_A(\langle |Q_B|^2\rangle-\bar Q_B^2)}.
\end{equation}
We now express $\lambda$ in terms of the resolution parameter,
Eq.~\eqref{defchiA}. Using the factorization property
Eq.~\eqref{spproduct}, we rewrite Eq.~\eqref{defchiA} as 
\begin{equation}
\label{defchiA2}
\frac{1}{\chi_A^2}+1\equiv
\frac{\langle |Q_{A}|^2\rangle}{\bar Q_{A}^2} 
\end{equation}
or, equivalently, 
\begin{equation}
\label{defchiA3}
\chi_A^2\equiv \frac{\bar Q_{A}^2}{\langle |Q_{A}|^2\rangle-\bar Q_{A}^2}.
\end{equation}
In the absence of flow fluctuations, the resolution parameter $\chi_A$
is the ratio of the underlying flow \eqref{vn} to the standard
deviation of the flow vector around the mean
flow~\cite{Ollitrault:1997di}. 
Using Eq.~\eqref{defchiA3}, 
Eq.~\eqref{lambda2} can be rewritten as 
\begin{equation}
\lambda=\frac{\bar Q_A\chi_B^2}{\bar Q_B\chi_A^2}, 
\end{equation}
Inserting this value of $\lambda$ into Eq.~\eqref{defcombined}, one
recovers Eq.~\eqref{optweight}.

\end{document}